\documentclass[a4paper,11pt]{article}
\pdfoutput=1 

\usepackage{jinstpub} 

\usepackage{siunitx}
\usepackage{multirow}
\notoc

\title{Filters and Redundancies: An Exploration of Novel Coherent Noise Filters for High Energy Physics}

\author[a, 1]{Felipe Costa,\note{Corresponding author.}}
\author[a]{Nicolas Guimarães,}
\author[a]{Guilherme Milani,}
\author[a]{Bruno Sanches,}
\author[b]{Irakli Mandjavidze,}
\author[b]{Damien Neyret}
\author[a]{and Wilhelmus Van Noije}

\affiliation[a]{Escola Politécnica da Universidade de São Paulo,\\
Avenida Prof. Luciano Gualberto, 158, trav. 3, 05508-010 São Paulo, Brazil}
\affiliation[b]{CEA IRFU, Université Paris-Saclay,\\route de Saclay, 91191 Gif sur Yvette Cedex, France}

\emailAdd{felipe.w.costa@usp.br}

\abstract{This work presents radiation-tolerant implementations for the SALSA front-end readout ASIC through redundancy methods applied to two median-finding algorithms designed for coherent noise suppression. Bit-wise Median Finder (BWMF) and Combinatorial Sum Median Finder (CSMF) were implemented in TSMC \SI{65}{\nano\meter} and evaluated in terms of area, power, and latency. Three redundancy techniques were applied in this work to compare their impact: simple TMR, full TMR, and temporal TMR (TTMR). The simple and full TMR approach was applied in both algorithms to establish comparisons and TTMR was applied to CSMF as an improvement. The results indicate that the BWMF achieves efficient performance in terms of area and power under the simple TMR scheme, but exhibits significantly higher power consumption when using the more robust full TMR approach. The TTMR technique, in turn, offers reliable fault tolerance while maintaining a feasible balance between area and power.}


\proceeding{Topical Workshop on Electronics for Particle Physics\\
  October 6–10, 2025\\
  Rethymno, Crete, Greece}

\begin{document}
\maketitle
\flushbottom

\section{Introduction}
\label{sec:Introduction}

\begin{figure}[htbp]
\centering
\includegraphics[width=.9\textwidth]{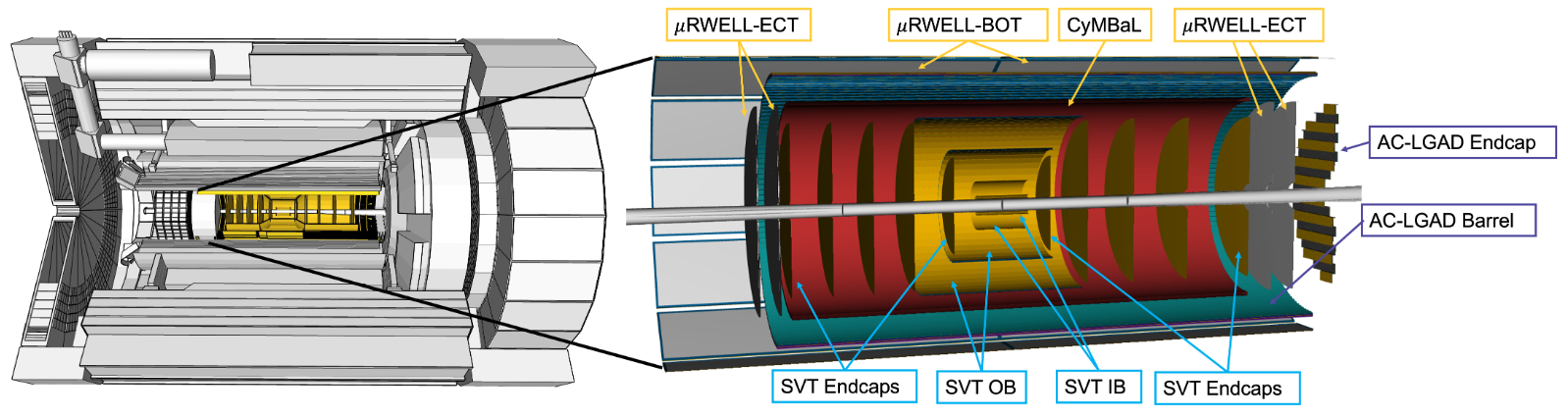}
\vspace{-.5cm}
\caption{\label{fig:Tracker} A view of the central region of the ePIC detector, showing a detailed 
schematic of the central tracker.}
\end{figure}

The ePIC (Electron-Proton/Ion Collider) detector is a general-purpose instrument designed to study the properties and behavior of subatomic particles such as gluons and quarks~\cite{ABDULKHALEK2022122447}, and it will be part of the EIC (Electron Ion Collider) accelerator at Brookhaven National Laboratory (BNL) in the USA~\cite{Accardi2016}. The ePIC detector includes a \SI{2}{\tesla} superconducting solenoid that provides sufficient magnetic rigidity for the reconstruction of secondary particle trajectories~\cite{11157721}. It also comprises layers of calorimeters and trackers; the latter is based on Monolithic Active Pixel Sensors (MAPS), Micropattern Gaseous Detectors (MPGD), and AC-LGAD (Capacitively-coupled Low-Gain Avalanche Diode) sensors (see Figure~\ref{fig:Tracker})~\cite{KUMAR2024169922}.

The MPGD detectors are implemented in both the inner and outer layers of the tracker barrel, using the $\mu$-RWELL (micro-Resistive WELL detector) and Micromegas technologies~\cite{SIDORETTI2025170622,AUNE2024169615,rafael2025cymbal,ACKER2020163423}. In the ePIC detector, the CyMBal employs Micromegas technology for the inner barrel layer, while the $\mu$-RWELL-ECT and $\mu$-RWELL-BOT are used in the end-cap and outer barrel trackers, respectively. Furthermore, the SALSA chip is a new readout ASIC designed to process the signals from all these MPGD detectors~\cite{Neyret_2025,gevin2025salsa}.

SALSA is an ASIC designed to read, amplify, and convert the analog signals from the MPGD detectors through an ADC~\cite{Sanches_2022, Hernandez, 8050227}. The chip includes 64 channels with 12-bit resolution each and is intended to operate at a sampling rate of \SI{50}{\mega Sps}~\cite{Sgroup2024}. It also integrates a Digital Signal Processor (DSP) that implements a chain of digital filters to process particle collision data, employing FIR, IIR, and slope-based filtering techniques~\cite{9307233, 9068974}. The DSP feeds four output links operating at a transmission rate of \SI{1}{\giga\bit/\second}, with an optional fifth link dedicated to the fast transmission of trigger information.

In the SALSA project, a median-finding approach is proposed to identify the median values read by the channels and apply them in a coherent noise-subtraction method to enhance data acquisition precision~\cite{10966346}. Considering the susceptibility of integrated circuits to Single-Event Effects (SEE) in radiation-prone environments like ePIC, this work evaluates the use of Triple Modular Redundancy (TMR) in the Bit-wise Median Finder (BWMF) and Combinatorial Sum Median Finder (CSMF) algorithms, focusing on area, power, and latency.

\section{Common Mode Noise Subtraction}
\label{sec:Common Mode Noise Subtraction}

\begin{figure}[htbp]
\centering
\includegraphics[width=.8\textwidth]{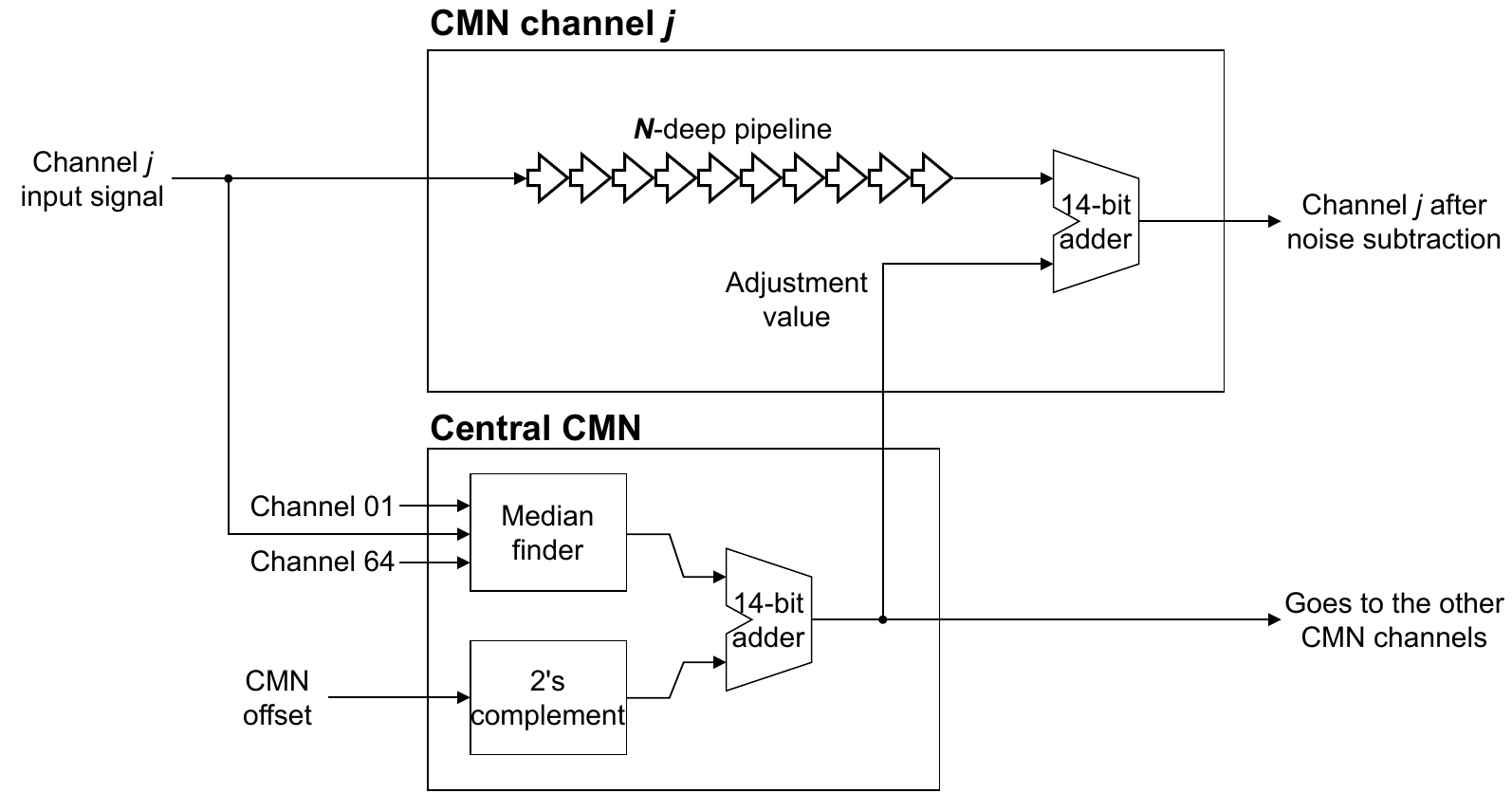}
\vspace{-.5cm}
\caption{\label{fig:CMN} The SALSA Common Mode Noise subtraction schematic.}
\end{figure}

The Common Mode Noise (CMN) subtraction (see Figure~\ref{fig:CMN})~\cite{Alme_2023} is a non-linear digital filtering technique designed to suppress coherent noise affecting multiple channels. At each sampling clock, the chip reads the 12-bit values from all channels. During a particle collision, however, only a few channels carry meaningful information, as most are not sufficiently excited by the detectors. In such cases, we propose that a median-finding filter can be used to evaluate the coherent noise and therefore mitigate its influence on the channels containing noise or secondary event signals. The median of the channel values serves as an indicator of coherent noise. Once this median is identified, it is used as a correction term subtracted from all channels to reduce the noise.

The CMN subtraction consists of two main blocks. The first is the Central CMN, which collects samples from all 64 channels and applies a configurable offset (CMN offset). The median value computed at this stage is then distributed to all CMN channel blocks. Each CMN channel features a deep pipeline that allows the Central CMN to calculate the median across all channels and apply the corresponding offset. Considering that the median-finding filter is the core element of the CMN subtraction process, we evaluate the performance trade-offs between two median-finding algorithms after the application of the TMR technique.

\section{Median-Finding Algorithms}
\label{sec:Median-Finding Algorithms}
One of the most important parts of the CMN that dictates the Power, Performance and Area (PPA) of the whole block is the median finding algorithm. In that case, the SALSA team developed two different algorithms for hardware to elaborate this block: one that relies more on a sequential approach called BWMF and another one that is based on combinatory logic named CSMF. 

\subsection{Bit-wise Median Finder (BWMF)}

The BWMF algorithm utilizes a pipeline composed of many stages in a way that each stage calculates one bit of the final median value. Therefore, this design achieves the final result in a number of clock cycles that is equal to the number of bits of resolution.

As our input is composed of 12 bits, the pipeline is also made up from 12 individual stages. Each of those is responsible for an iteration of the median value computation, in other words, the first stage evaluates the MSB of the median value and the last stage is responsible for the LSB.

The process of one single is represented by the Figure~\ref{fig:BWMFandCSMF} (left) and goes as the following: first, the algorithm analyzes the current bit position of every channel storing them in a 64-bit register, the value who appears more times (1 or 0) in this intermediate storage will be the partial median bit for this position. Now it is necessary to update the values of the candidates (channel values), that is done by comparing its bit for the current position with the current partial median bit, if they are different the channel is no more a candidate and all of its bits are replaced by 0's if its bit for the current position is 0 and by 1's if otherwise. The channels that have the same bit as the partial median are kept without changes.

This process repeats itself for the 12 stages, and when that finished the partial median value becomes the true median value of those channels, and it is propagated to the output.

The sequential nature of this algorithm, enables a small number of elements, enabling a design that has both low power consumption and a small area. However, it is a high latency module that has many registers in its composition, hindering the implementation of radiation hardening techniques. 

\subsection{Combinatorial Sum Median Finder (CSMF)}

The CSMF approach goes in the other direction, leveraging a highly combinatorial structure to enable the maximum number of parallelized computations. This characteristic allows the result to be achieved in a single clock cycle.

\begin{figure}[htbp]
\centering
\includegraphics[width=.47\textwidth]{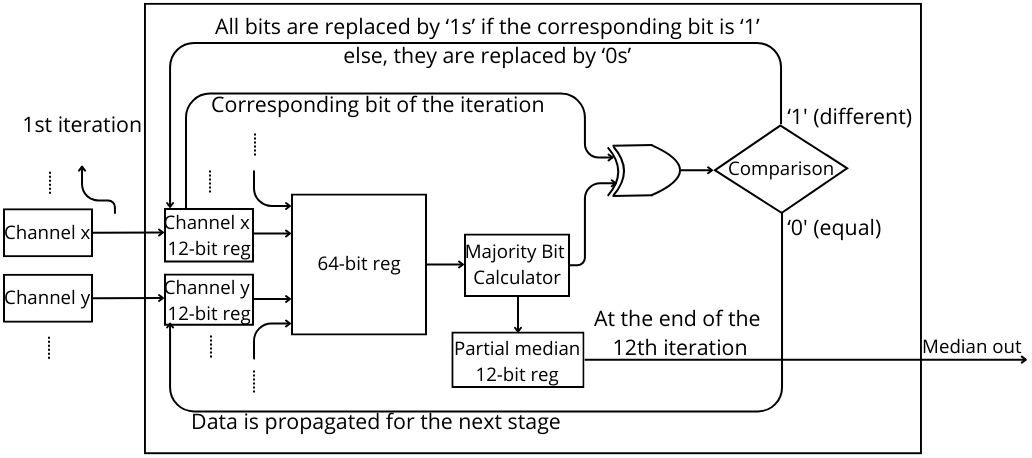}
\qquad
\includegraphics[width=.47\textwidth]{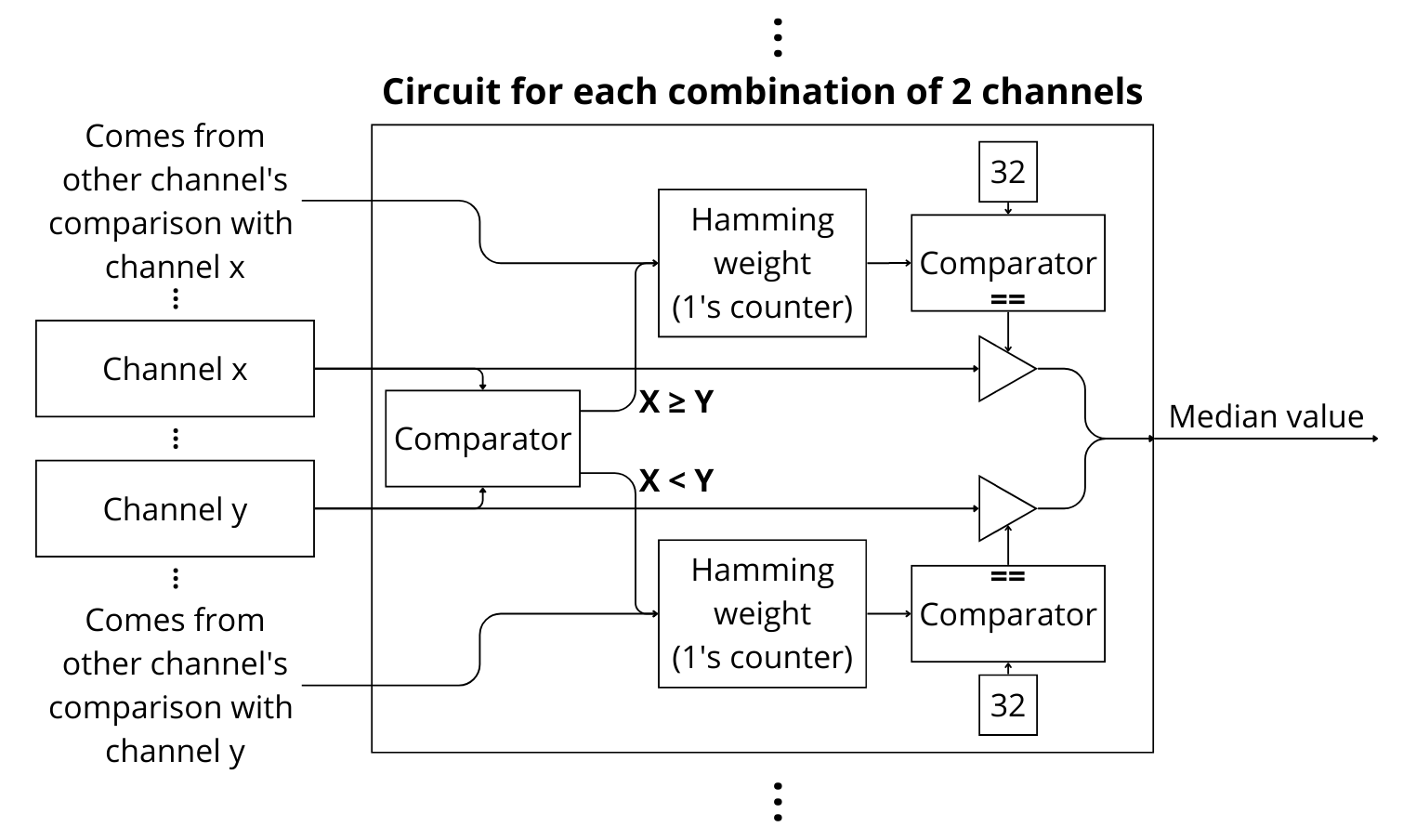}
\vspace{-.75cm}
\caption{\label{fig:BWMFandCSMF} BWMF (left) and CSMF (right) algorithm diagram showing the data path.}
\end{figure}

Its architecture, demonstrated in Figure~\ref{fig:BWMFandCSMF} (right), is composed of 2016 comparators, one for each unique pair of channels: $\binom{64}{2} = \frac{64!}{2! \, (64 - 2)!}  = 2016$

Apart from the comparators, each channel has a hamming weight block (a 1's counter). Each comparison of a channel X with a channel Y, sends a 1 signal to the hamming weight of the greater value and a 0 to hamming weight of the smaller channel. Therefore, the hamming weight will count how many channels are smaller than its respective channel and the median value can be easily found by taking the channel whose 1's counter has the value of 31, in other words the 32nd greatest value.

It is important to clarify that the correct median value would be calculated from the mean value between the 32nd and the 33rd greatest values, but here the role of this modules is to quantify a value for the common mode noise. With that in mind, a simple approximation is sufficient.

Considering the structure of this module, its highly parallelized nature makes it  bigger and more power hungry when compared to the BWMF. But it has some highlights: the low latency and small number of sequential elements which facilitates the implementation of radiation hardening techniques. 

\section{Triple Modular Redundancy and Single-Event Effects}
\label{sec:Triple Modular Redundancy and Single-Event Effects}

The ideas for further TMR were first introduced by von Neumann~\cite{von1956probabilistic}. The fundamental principle of TMR is to replicate the logic three times and employ a majority voter to determine the correct output. This technique is used to mitigate logic errors induced by charged particles, referred to as SEE, which can be further classified into soft errors and hard errors. Several techniques exist to protect integrated circuits against SEE, most of which rely on data redundancy. When a bit of information is stored across multiple nodes, it is possible to reconstruct the original data even if some of these nodes are affected by transient disturbances.

An SEE occurs when a charged particle strikes a silicon device, ionizing the material and generating electron–hole pairs~\cite{Aguiar2025}. The charge deposited by the particle may be collected by the drain or source diffusion region, potentially altering the logic state. Non-destructive SEE (soft errors) are typically classified into two main types: Single-Event Transients (SET) and Single-Event Upsets (SEU).

An SET is a phenomenon occurring in combinatorial logic, manifesting as a short-lived glitch on a net that lasts for a few nanoseconds until the correct logic value is restored. The likelihood of capturing an SET increases with higher clock frequencies, as the clock period becomes comparable to the SET duration~\cite{659037,6530775}. In contrast, an SEU occurs in memory elements such as flip-flops or latches, where the affected cell cannot restore its previous value~\cite{819104}.

\section{Comparison}
\label{sec:Comparison}

\begin{figure}[htbp]
\centering
\includegraphics[width=.39\textwidth]{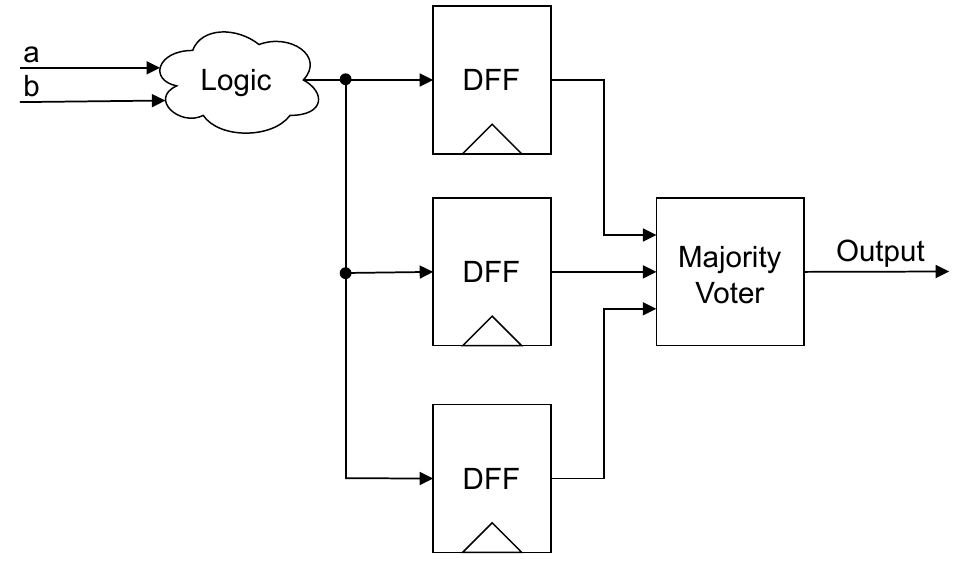}
\qquad
\includegraphics[width=.39\textwidth]{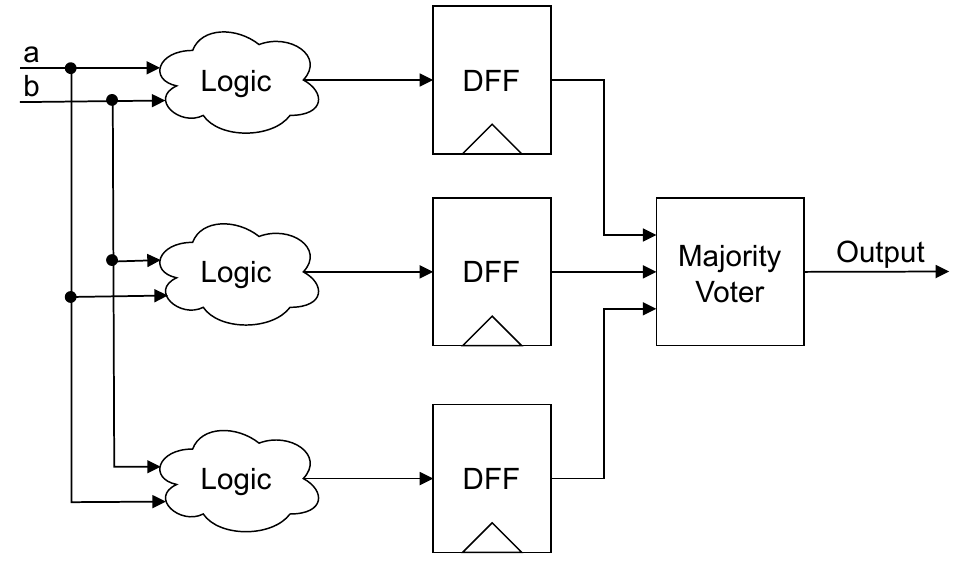}
\qquad
\includegraphics[width=.39\textwidth]{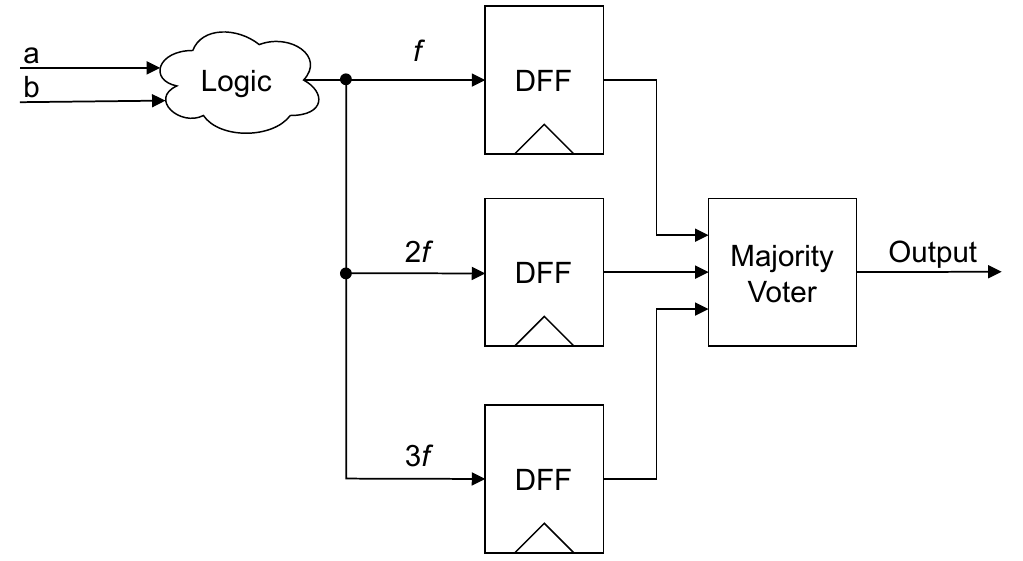}
\vspace{-.25cm}
\caption{The top-left schematic illustrates the simple TMR approach, the top-right schematic illustrates the full TMR implementation and the bottom schematic illustrates the TTMR approach.\label{fig:TMRschematics}}
\end{figure}

For the median-finding filters presented in Section~\ref{sec:Median-Finding Algorithms}, three implementation approaches are proposed to protect the BWMF and CSMF filters against ionizing particle strikes (see Section~\ref{sec:Triple Modular Redundancy and Single-Event Effects}): filters with no TMR, filters with simple TMR (where only the memory cells are triplicated), and filters with full TMR, in which both logic and memory cells are replicated. In addition, since the CSMF is mainly a combinational design, the Temporal TMR (TTMR) approach is introduced as an enhancement, as this architecture is particularly sensitive to SETs (see Figure~\ref{fig:TMRschematics}), where the logic value is stored in different memory elements operating at multiples of the base clock frequency.

For logic synthesis and physical implementation, the Cadence design flow was adopted, using the Genus and Innovus tools to obtain the performance results. The main steps for logical and physical synthesis with the TMR technique can be summarized as follows: the process begins with a Verilog or SystemVerilog design description; the TMR technique is then applied, and in the case of full TMR, the TMRG tool~\cite{Kulis2017} is employed to apply the TMR automatically. For logic synthesis, the TSMC \SI{65}{\nano\meter} standard-cell library is used, along with a constraint file that prevents the synthesis tool from optimizing or merging the triplicated memory cells. Finally, during physical implementation, an additional constraint file is provided to ensure that the triplicated macro cells are placed at a safe distance from each other (a spacing of 15 placement units is adopted in this work). The results of the BWMF and CSMF is summarized on Table~\ref{tab:TMRcomparison}.

\begin{table}[htbp]
\centering
\caption{\label{tab:TMRcomparison} Comparison results for the BWMF and CSMF filters.}
\smallskip
\begin{tabular}{|c|c|c|c|c|c|}
\hline
Filter & Redundancy & No TMR & Simple TMR & Full TMR & TTMR \\
\hline
\multirow{3}{*}{BWMF} 
 & Number of Cycles & 10 & 10 & 10 & --- \\
 & Area $\left[\text{mm}^2\right]$ & 0.039 & 0.039 & 0.128 & --- \\
 & Power $\left[\text{mW}\right]$ & 3.50 & 6.95 & 105.48 & --- \\
\hline
\multirow{3}{*}{CSMF} 
 & Number of Cycles & 1 & 1 & 1 & 1 \\
 & Area $\left[\text{mm}^2\right]$ & 0.133 & 0.141 & 0.423 & 0.145 \\
 & Power $\left[\text{mW}\right]$ & 8.48 & 8.37 & 24.77 & 11.26 \\
\hline
\end{tabular}
\end{table}

The BWMF is a sequential implementation of an innovative median-finding algorithm, contains a larger number of flip-flops than the CSMF, requiring the triplication of memory cells for SEU protection and thereby increasing both area and power consumption. In contrast, the CSMF is an almost fully combinatorial design and is more susceptible to SETs. The CSMF shows worse area performance for both simple and full TMR, but the BWMF consumes about $4.26\times$ more power under full TMR, which is a clear disadvantage. When the latency of both algorithms is compared, the CSMF also overcomes the BWMF, since the latter takes $10\times$ more cycles to finish one computation. The TTMR approach proposed for the CSMF emerges as the optimal solution since it uses both temporal and spatial redundancy, reducing both area and power, despite introducing additional design complexity.

\section{Conclusion}
\label{sec:Conclusion}

In this work, three TMR approaches are evaluated for the median-finding filters in the CMN stage of the SALSA chip. In the radiation-prone environment of the ePIC detector, understanding their trade-offs is crucial. The CSMF, which employs the TTMR approach, achieves significantly lower power dissipation with only a slight area increase compared to the full TMR implementation of the BWMF, while still providing a reliable TMR technique, making it a feasible choice for the final design.

\acknowledgments

This work was financed, in part, by the São Paulo Research Foundation (FAPESP), Brazil, grants \#2024/04802-9, \#2024/06703-8, CNPQ Grant \#134869/2024-9, grant number ANR-24-CE31-7003-01. This study was financed in part by the Coordenação de Aperfeiçoamento de Pessoal de Nível Superior – Brasil (CAPES) – Finance Code 001. The authors also thank CERN and Kulis et al. for making available the TMRG tool that was used in this work.


\bibliographystyle{JHEP.bst}
\bibliography{referencesabbrev}
\end{document}